\documentclass[conference,a4paper,10 pt]{IEEEtran}
\IEEEoverridecommandlockouts

\usepackage[hidelinks]{hyperref}
\usepackage[cmex10]{amsmath}
\usepackage{amssymb,amsfonts}
\usepackage{dblfloatfix}

\usepackage[ruled,vlined]{algorithm2e}
\usepackage{graphicx}
\graphicspath{{Figures/PDF/}{Figures/PNG/}}

\usepackage{booktabs}
\usepackage{siunitx}
\usepackage[numbers,compress]{natbib}
\usepackage{texnames}
\usepackage{bm,bbm}
\usepackage{orcidlink}

\begin{document}

\title{\uppercase{ From very fine Vertical Resolution to Range Super-Resolution on BIOMASS P-BAND SAR Data}
\thanks{This work has been submitted to the IEEE for possible publication. Copyright may be transferred without notice, after which this version may no longer be accessible.}

}


	
    
    
	
    

\author{%
  \IEEEauthorblockN{%
    Yanghai Yu\orcidlink{0000-0001-9227-1091}\IEEEauthorrefmark{1}\IEEEauthorrefmark{2},
    Stefano Tebaldini\orcidlink{0000-0002-1229-3811}\IEEEauthorrefmark{1},
    Yang Lei\orcidlink{0000-0002-8377-1980}\IEEEauthorrefmark{2},
    and Fabio Rocca\orcidlink{0000-0002-2266-6212}\IEEEauthorrefmark{1}}
  \IEEEauthorblockA{%
    \IEEEauthorrefmark{1}\textit{RIDE Lab, Dipartimento di Elettronica, Informazione e Bioingegneria, Politecnico di Milano}, 20133 Milan, Italy\\
    \IEEEauthorrefmark{2}\textit{ National Space Science Center, Chinese Academy of Sciences}, 100190 Beijing, China}
}

\maketitle
\begin{abstract}
	This paper demonstrates range super-resolution over arid regions using SAR tomography data with very fine vertical resolution from the European Space Agency's BIOMASS mission. As the first civilian spaceborne P-band SAR mission, BIOMASS is operating with a bandwidth limited to 6~MHz by International Telecommunication Union regulations, yielding a native slant-range resolution of approximately 25~m.  The large spatial baseline data acquired during the commissioning phase in July and August 2025 provide a valuable opportunity to enhance range resolution based on the InSAR wavenumber shift principle. According to this principle, a wider effective bandwidth can be synthesized by combining complementary sub-bands of the ground reflectivity spectrum sampled across acquisitions at different look angles. Unlike existing literature, this paper offers a novel interpretation of this problem from the perspective of SAR tomography. Furthermore, the proposed method also accounts for residual topography errors. The range resolution enhancement performance is validated using all 20 available acquisitions from the July–August 2025 commissioning phase over Lop Nur, China, and Wadi al-Hayat, Libya, achieving spatial detail comparable to  optical satellite imagery in a grid finer than 2 m. Furthermore, the reconstructed range spectrum of the central Libyan oasis exhibits an effective bandwidth exceeding 70 MHz. The point spread function (PSF) of a point like target in the scene demonstrates a slant-range resolution finer than 2.1 m, representing more than a twelvefold improvement over the native resolution.
\end{abstract}


\section{Introduction}

Spatial resolution characterizes an imaging system's ability to distinguish closely spaced features and serves as a fundamental metric of image quality in remote sensing. While many Earth observation applications rely on fine spatial detail, sensor design and hardware constraints often limit the resolution attainable during acquisition. To overcome these limitations, super-resolution (SR) methods have been developed to reconstruct higher-resolution images from lower-resolution observations~\cite{SR_review}.

A compelling example illustrating the need for SR techniques is the European Space Agency's (ESA) BIOMASS mission. As the first civilian spaceborne P-band (435~MHz) Synthetic Aperture Radar (SAR) mission, BIOMASS is primarily designed to generate near-global maps of forest above-ground biomass (AGB) at nine-month intervals~\cite{QUEGAN201944, scipal2026biomass}. Besides its primary  objective, the mission also supports several secondary applications, including subsurface imaging in cryospheric \cite{banda2026first,ICE_DLR} and arid environments \cite{Philippe_desert,yu2026preliminary} and ionospheric structure characterization \cite{tebaldini2026multi}. However, International Telecommunication Union (ITU) regulations limit the radar bandwidth to approximately 6~MHz, yielding a native slant-range resolution of about 25~m. In particular, the large spatial baseline data acquired in July and August, 2025 within the commissioning phase allows the improvement of range resolution based on the InSAR wavenumber shift theory \cite{gatelli1994wavenumber,prati_spectrum}. 

According to InSAR wavenumber-shift theory, acquisitions at different look angles sample shifted portions of the ground reflectivity spectrum. Although the resulting spectral shifts reduce spectral overlap, the corresponding subband signals can be coherently combined to extend the effective range bandwidth of the ground spectrum, thereby improving range resolution. Previous studies have demonstrated this improvement
using real SAR data~\cite{prati_spectrum,fornaro2006_rg_enh,airborne_rg_res_enh}.  Unlike existing literature, this work interprets InSAR wavenumber shift principles from a new tomographic perspective, enabling simultaneous range resolution enhancement and topographical error estimation.

High temporal stability across passes is another critical requirement for range resolution enhancement in BIOMASS repeat-pass acquisitions. To meet this criterion, we selected two arid test sites for our experiments: a high-salinity desert region in China and a hyper-arid region in central Libya. We processed all 20 passes acquired between July and August 2025 to evaluate the optimal performance for range resolution enhancement. Comparisons against high-resolution optical imagery, alongside analyses of the Point Spread Function (PSF) and the reconstructed effective range spectrum, demonstrate  an achieved range resolution of approximately 2 m.

Section~\ref{sec_inter} interprets InSAR wavenumber shift theory using SAR tomography. Section~\ref{sec_method} describes the processing flow, and Section~\ref{sec_study_site} introduces the study sites and BIOMASS data. Sections~\ref{sec_res} and~\ref{sec_con} present the results and conclusions, respectively.

\section{Interpreting InSAR Wavenumber Shift Theory Through SAR Tomography}\label{sec_inter}

\subsection{Range and Vertical Resolution}

For a single synthetic aperture radar (SAR) image with processed range bandwidth
$B$, the nominal slant-range resolution is $\rho_r = c/(2B)$, where $c$ is the
speed of light. Combining SAR images acquired at different look angles
synthesizes an elevation aperture. The resulting vertical resolution is
$\rho_z = 2\pi/\Delta\kappa_z$, where $\Delta\kappa_z$ denotes the maximum
magnitude of the interferometric vertical wavenumber $\kappa_z$ over all
acquisition pairs. Following~\cite{bamler1998synthetic}, we express $\kappa_z$
for two acquisitions whose look angles differ by $\delta\theta$ as
\begin{equation}
    \kappa_z = \frac{4\pi}{\lambda}\frac{\delta\theta}
                    { \sin(\theta_{\mathrm{ref}} - \alpha)},
\end{equation}
where $\lambda$ is the radar wavelength, $\theta_{\mathrm{ref}}$ is the
reference look angle, and $\alpha$ is the local terrain slope angle.

\subsection{InSAR Wavenumber Shifts in Multibaseline Acquisitions}
\begin{figure}
    \centering
    \includegraphics[width=0.9\linewidth]{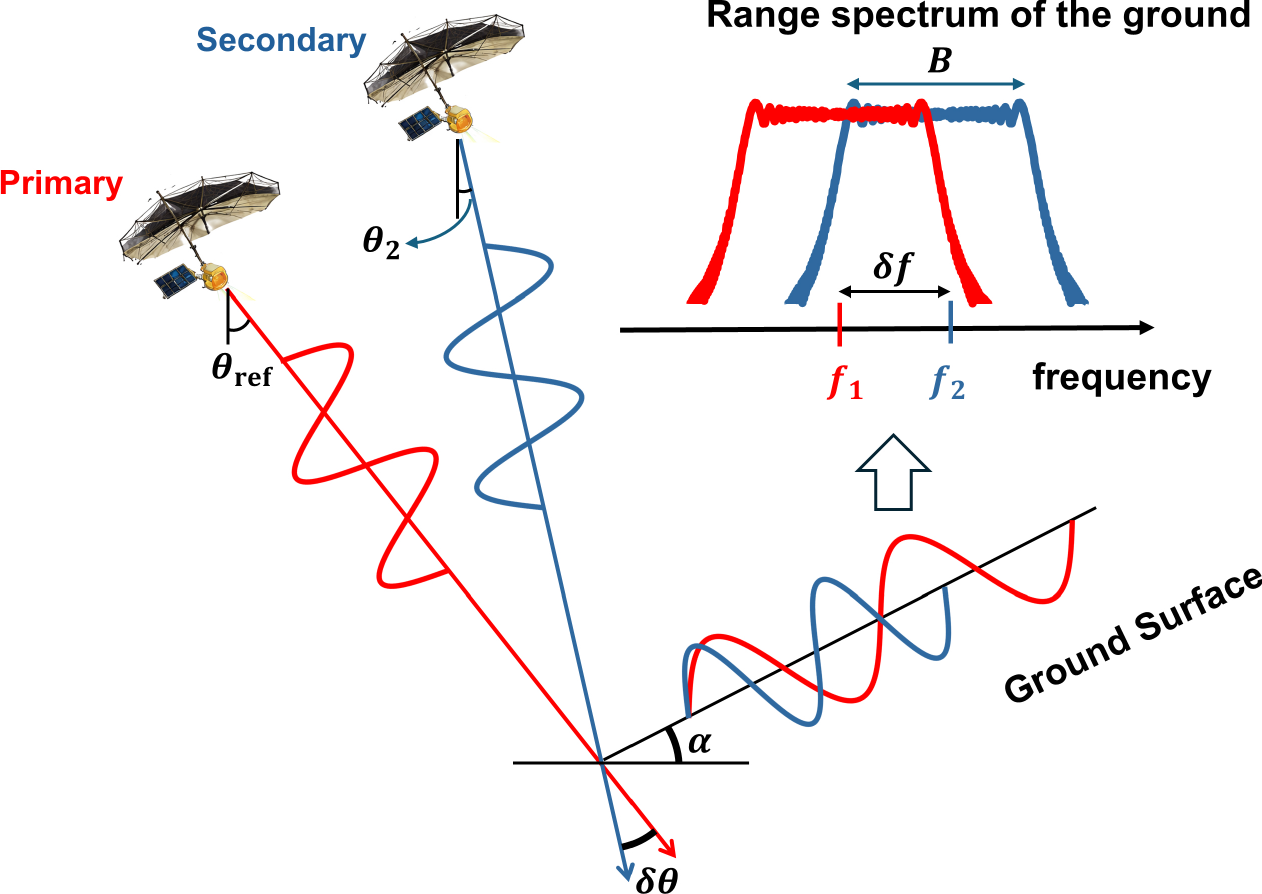}
    \caption{Schematic illustration of the InSAR wavenumber shift theory for acquisitions at two different look angles.}
    \label{fig_insar_shift_wavenumber_illu}
\end{figure}

 InSAR wavenumber shift concept describes how SAR observations acquired at different look angles sample different portions of the ground reflectivity spectrum (see Fig.~\ref{fig_insar_shift_wavenumber_illu}). Specifically, for a single horizontal layer, a look-angle separation $\delta \theta$ between two SAR acquisitions induces a frequency shift $\delta f$, given by
\begin{equation}\label{eq_insar_freq_shif}
    \delta f = -\frac{\delta \theta}{\tan(\theta_{\mathrm{ref}}-\alpha)},
\end{equation}

This expands the effective range bandwidth to $B' = B + \lvert\delta f\rvert$. Continuous wideband coverage requires multibaseline SAR acquisitions with sufficiently large spatial baseline separations while maintaining spectral overlap between spatially adjacent acquisitions. This overlap facilitates topography retrieval and residual phase calibration. The enhanced range resolution is therefore theoretically given by:
\begin{equation}\label{rg_res_new}
    \rho'_r = \frac{c}{2B'}=\frac{\rho_r\rho_z}{\rho_r\cos\left(\theta_{\mathrm{ref}}-\alpha\right)+\rho_z},
\end{equation}

\subsection{Insights from SAR tomography}

\begin{figure}
    \centering
    \includegraphics[width=0.9\linewidth]{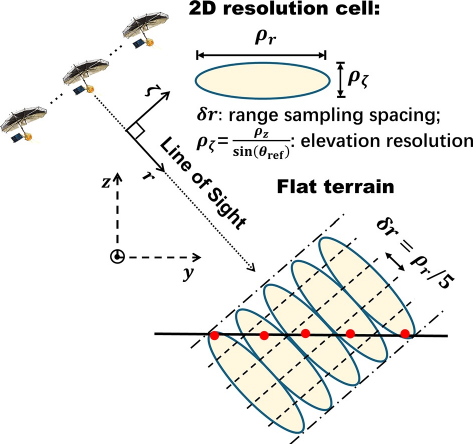}
    \caption{Schematic illustration of range resolution enhancement over flat terrain from a tomographic perspective. Here, $y$, $z$, and $\zeta$ denote the ground range, vertical, and elevation directions, respectively. }
    \label{fig_interp_tomosar}
\end{figure}

This subsection provides a tomographic interpretation of the relationship \eqref{rg_res_new}. Fig.~\ref{fig_interp_tomosar} illustrates the BIOMASS large-baseline acquisition geometry, where vertical resolution is roughly one-fifth of range resolution. For a flat surface sampled at a coarse interval $\delta r$ comparable to range resolution, scatterers within a single interval (red dots) map to distinct apparent range–elevation cells. This represents that tomographic acquisitions with fine vertical resolution can therefore distinguish horizontally separated targets in different range bins.

\section{Phase calibration and reconstruction}\label{sec_method}
This paper adopts a straightforward calibration and topography estimation strategy. Following ionospheric calibration~\cite{tebaldini2026multi}, constant phase offsets are initially estimated for each image on the multilooked radar grid (e.g., using a $100 \times 100\text{ m}^2$ window) based on the procedure documented in \cite{yu2026preliminary}. Subsequently, SAR tomography-based methods \cite{tebaldini2015phase} are applied for phase calibration and topography estimation. To reconstruct the wideband signal, the sub-band summation method \cite{prati_spectrum} is employed due to its simplicity and computational efficiency compared with the matrix-based inversion method proposed in \cite{fornaro2006_rg_enh}. This approach involves upsampling the baseband BIOMASS signals and applying a phase rotation that accounts for distance information.

\section{Study Sites and BIOMASS data}\label{sec_study_site}
This study evaluates two arid test sites. The first, in the \textit{Fezzan region of Libya}, covers a village in Wadi al-Ajal, an oasis between the Ubari Sand Sea and the Messak Settafet escarpment. Groundwater supports irrigated fields, date palm groves, and settlements amid hyper-arid desert. The second, in the \textit{Lop Nur region of China}, covers part of a potash plant in the eastern Tarim Basin, Xinjiang, where subsurface brine is extracted for fertilizer production.

We perform range resolution enhancement using 20 SAR images per site acquired in July--August 2025. For these acquisitions, baseline separation is greatest near the equator and decreases toward the poles, yielding vertical resolutions of approximately 1.9~m at the Libyan oasis site ($26.54^\circ\mathrm{N}$) and 2.2~m at the Lop Nur site ($40.44^\circ\mathrm{N}$). 

\section{Experimental Results}\label{sec_res}

\begin{figure}
    \centering
    \includegraphics[width=0.9\linewidth]{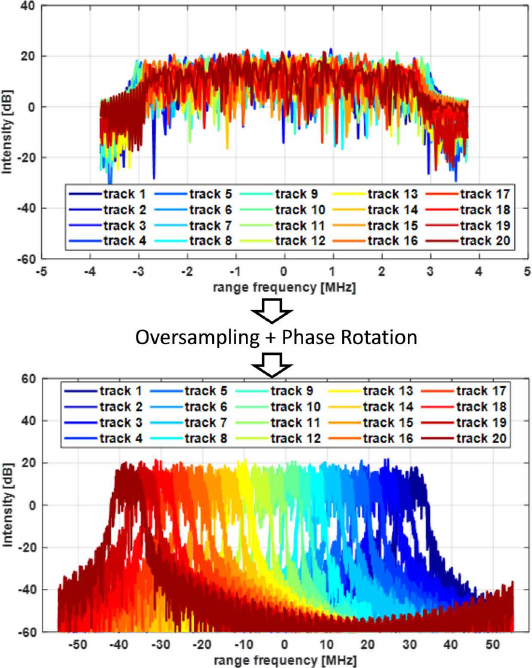}
    \caption{Range spectra of 20 BIOMASS acquisitions along a single range line over the Libya study site: (top) baseband signals and (bottom) passband signals after oversampling and phase rotation. }
    \label{fig_spectral_shift_ill}
\end{figure}

\begin{figure}
    \centering
   \includegraphics[width=1\linewidth, trim=0cm 0cm 0cm 0cm, 
  clip=true]{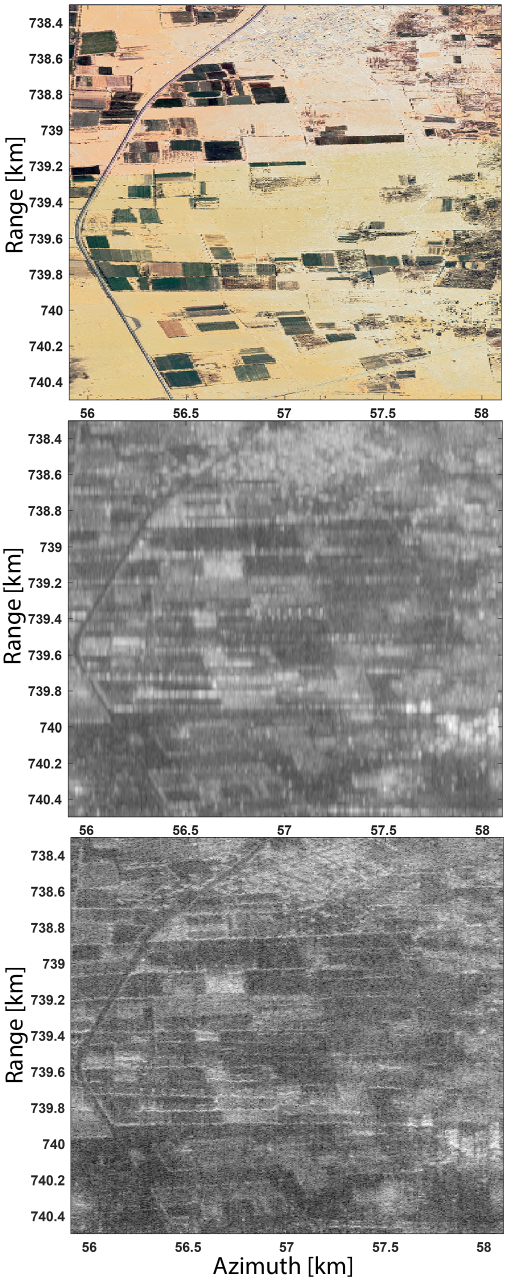}
    \caption{Optical and SAR images of an oasis in central Libya ($26.54^\circ\mathrm{N}$, $13.38^\circ\mathrm{E}$) in radar grid. (Top) High-resolution optical image from Google Earth sampled on a grid with 1.8~m spacing. (Middle) Incoherent average of 20 BIOMASS images. (Bottom) Range super-resolution reconstruction from 20 BIOMASS images. }
    \label{fig_libya_comp}
\end{figure}

\begin{figure}
    \centering
    \includegraphics[width=1\linewidth,trim=0cm 0cm 0cm 0cm, 
  clip=true]{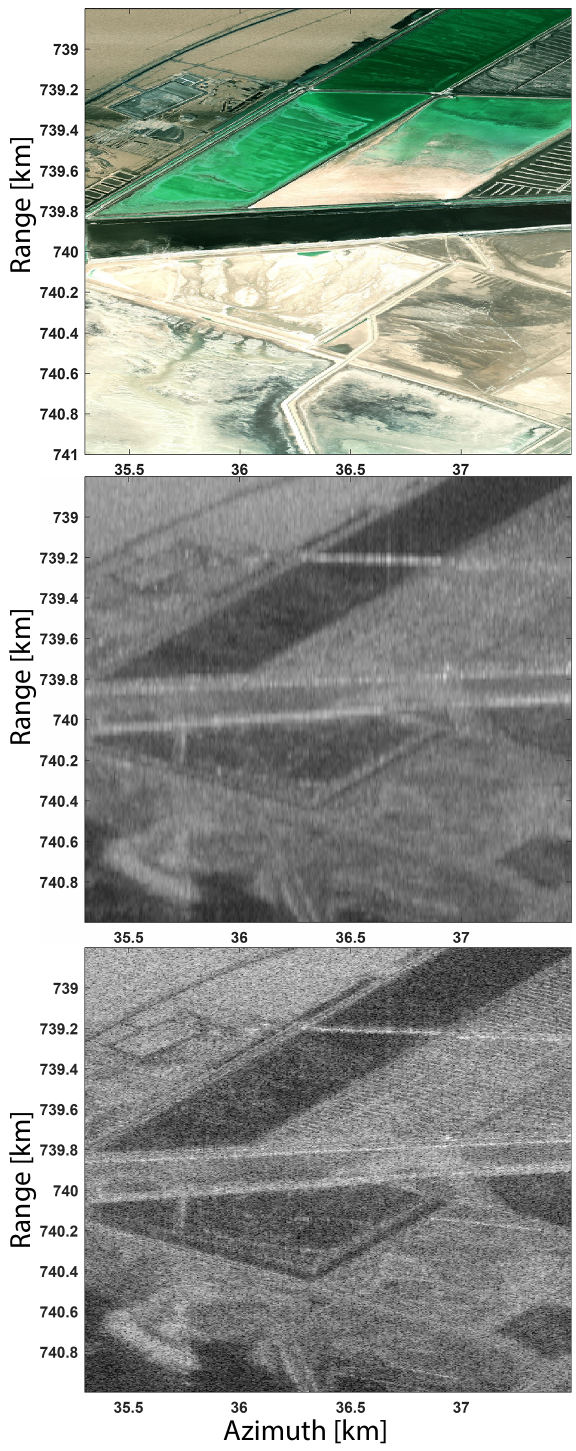}
    \caption{Optical and SAR images of a potash plant in Xinjiang, China ($40.46^\circ\mathrm{N}$, $90.86^\circ\mathrm{E}$) in radar grid. (Top) High-resolution optical image from ESRI satellite map sampled on a 1.8~m grid. (Middle) Incoherent average of 20 BIOMASS images. (Bottom) Range super-resolution reconstruction from 20 BIOMASS images. }
    \label{fig_lopnur_comp}
\end{figure}

\begin{figure}
    \centering
    \includegraphics[width=0.95\linewidth]{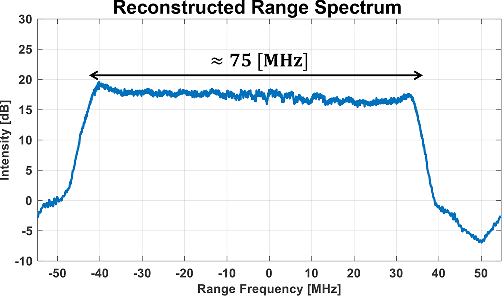}
    \caption{Range spectrum of the reconstructed scene at the Libyan site.}
    \label{fig_real_final_range_spectrum}
\end{figure}

\begin{figure}
    \centering
    \includegraphics[width=0.8\linewidth]{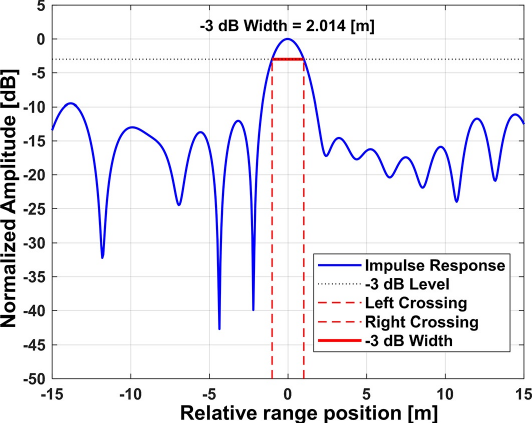}
    \caption{PSF of a point-like target at the Libyan site.}
    \label{fig_range_psf}
\end{figure}

This section first analyzes the spectral behavior of 20 BIOMASS images, then compares the enhanced-resolution results against high-resolution optical imagery and the original BIOMASS images, and finally evaluates the effective range spectrum and point spread function (PSF) of a point-like target.

The upper panel of Fig.~\ref{fig_spectral_shift_ill} shows the aligned 6~MHz baseband spectra of a single range line from the 20 BIOMASS acquisitions at the Libyan site. The data are oversampled using a factor based on 80\% of the range resolution predicted by \eqref{rg_res_new} and phase-rotated using the range estimates from Section \ref{sec_method}. The shifted spectra collectively span a continuous band up to 75~MHz wide. The approximately 3~MHz separation between consecutive acquisitions suggests that correlation is generally limited to spatially adjacent pairs.

Based on the processing in Section~\ref{sec_method}, we generated range super-resolution images of the Libyan oasis and Lop Nur potash plant regions (see Figs.~\ref{fig_libya_comp} and~\ref{fig_lopnur_comp}, respectively). We compare these with Google Earth optical imagery sampled on a 1.8~m grid and incoherent averages of the 20 SAR acquisitions, which offer high radiometric quality at 25~m range resolution. The enhanced images resolve details comparable to the optical reference and reveal additional features in the upper part of the village and its surroundings. Sandy-area responses may indicate buried features. At Lop Nur site, fine textures are both visible in the optical and super-resolution SAR images while remain unresolved in the incoherent averaged imagery.

Finally, the range spectrum of the reconstructed scene (Fig.~\ref{fig_real_final_range_spectrum}) and the PSF of a point-like target along range direction (Fig.~\ref{fig_range_psf}) indicate a range resolution of approximately 2.01~m, based on the $-3$~dB PSF width.

\section{conclusion}\label{sec_con}
This work demonstrates range super-resolution using BIOMASS data in arid environments, achieving a more than twelvefold improvement over the native range resolution. Future work will systematically assess the effect of the temporal stability and explore a broader range of applications. Further, while the wavenumber shift approach considered only flat ground acquisitions, the tomographic one will allow a full physical understanding of the targets, combining surface and volume characterizations.

\small
\bibliographystyle{IEEEtranN}
\bibliography{references}

\end{document}